\documentclass[%
 reprint,
 amsmath,amssymb,
 aps,
]{revtex4-2}

\usepackage{graphicx}
\usepackage{dcolumn}
\usepackage{bm}
\usepackage{siunitx}
\usepackage{xspace}

\usepackage{xcolor}
\usepackage{soul}
\usepackage[utf8]{inputenc}
\usepackage{lineno}
\usepackage{float}

\newcommand{\refSensitivity}{Figure~\ref{fig:sensitivity}\xspace}

\newcommand{\meff}{m_{\textrm{eff}}}
\newcommand{\gammam}{\gamma_{\textrm{m}}}
\newcommand{\gammag}{\gamma_{\textrm{g}}}
\newcommand{\gammasc}{\gamma_{\textrm{sc}}}
\newcommand{\gammatot}{\gamma_{\textrm{tot}}}
\newcommand{\Qm}{Q_{\textrm{m}}}
\newcommand*\diff{\mathop{}\!\mathrm{d}}
\newcommand{\Fdisk}{\mathcal{F}_{\textrm{mem}}}
\newcommand{\Fcavity}{\mathcal{F}_{\textrm{cav}}}
\newcommand{\Mgas}{M_{\textrm{gas}}}

\newcommand{\Qfactor}{$Q$-factor\xspace}
\newcommand{\Fopt}{F_{\textrm{opt}}}
\newcommand{\kopt}{k_{\textrm{opt}}}
\newcommand{\Ptrap}{\mathcal{P}_\mathrm{trap}}
\newcommand{\omegaDM}{\omega_{\textrm{DM}}}

\newcommand{\mDM}{m_{\textrm{DM}}}

\begin{document}



\title{An optomechanical platform for high-frequency gravitational wave and \\ vector dark matter detection}

\author{David Rousso}
\email{david.rousso@cern.ch}
\affiliation{Deutsches Elektronen-Synchrotron DESY, 22603 Hamburg, Germany}

\author{Moritz Bjoern Kristiansson Kunze}
\email{moritz.kunze@online.de}
\affiliation{Deutsches Elektronen-Synchrotron DESY, 22603 Hamburg, Germany}

\author{Christoph Reinhardt}
\email{christoph.reinhardt@desy.de}
\affiliation{Deutsches Elektronen-Synchrotron DESY, 22603 Hamburg, Germany}


\date{\today}

\begin{abstract}
We propose the Nano-membrane Experiment for Space-time Tremors (NEST), a nanomechanical membrane as input mirror of a 100~m optical cavity, to detect high-frequency gravitational waves and vector dark matter.
Cavity-length modulation resonantly drives the membrane, whose frequency is tuned \textit{in situ} by optical trapping.
In a limiting best-case scenario, NEST covers 0.5--40~kHz with a peak strain sensitivity of $3\times 10^{-23}/\sqrt{\mathrm{Hz}}$, toward which six membranes are proposed as a first implementation.
A germanium reference mass adds sensitivity to $B\!-\!L$ dark matter, in the best case beyond existing limits.
\end{abstract}

\maketitle

\section{Introduction and Detector Concept}
Ten years since the advent of gravitational wave (GW) detection \cite{abbott2016observation}, the  LIGO, Virgo, and KAGRA observatories have enabled new tests of general relativity and a deeper understanding of black holes and neutron stars \cite{bailes2021gravitational}. 
These kilometer-scale interferometers are optimized for frequencies up to a few kHz, beyond which their sensitivity becomes limited by shot noise \cite{capote2025advanced,schnabel2025optical}. 
Toward the upper end of their observational band, an especially enticing target are continuous GW signals generated by deformations of rapidly rotating neutron stars in the 700–\SI{1500}{Hz} range \cite{NeutronStarCGWs}, as well as ringdown GW signals from neutron star mergers in the 2--\SI{4}{kHz} range \cite{NeutronStarRingdownGWs}, both of which can probe the equation of state of ultra-dense matter and help explain the observed paucity of fast pulsars.
Furthermore, several beyond-Standard-Model (BSM) scenarios predict GW signals at even higher frequencies beyond the kHz-range, including those from primordial black hole mergers, axion superradiance, cosmic strings, and early-universe phase transitions \cite{HFGW,sprague2024simulating}. 
This has fueled the development of novel high-frequency gravitational wave (HFGW) detectors that complement large-scale laser observatories \cite{HFGW,patra2025broadband}.

Current approaches include detectors based on superconducting cavities \cite{berlin2022detecting,fischer2025first}, circuits \cite{domcke2022novel}, and magnets \cite{domcke2025magnets}; broadband electromagnetic detectors \cite{capdevilla2025high}, probes of the frequency or polarization of laser light \cite{bringmann2023high,garcia2025polarimetric}; bulk acoustic wave resonators \cite{goryachev2014gravitational,goryachev2021rare,linehan2025listening}, and superfluid helium \cite{vadakkumbatt2021prototype}. 
For example, the bulk acoustic wave experiment MAGE, operating at 5--\SI{15}{MHz}, has excluded mergers of planetary-mass primordial black holes (PBHs) of \SI{1.7e-9}{}--\SI{1.2e-4}{M_{\odot}} occurring more often than once per year within \SI{1e-6}{kpc} \cite{campbell2025experimental}.
Another concept is the levitated sensor detector (LSD), in which the mechanical resonance of an optically-levitated nanodisk inside a 100-m optical cavity transduces the GW strain into an optical signal \cite{Nanodisk2013,Nanodisk2022}. 
While theoretical sensitivity to BSM signatures has been demonstrated, the LSD requires maintaining stable levitation of a high-aspect-ratio particle in ultra-high vacuum, and a cavity end mirror of radius $\sim\SI{1}{m}$ to suppress scattering at the nanodisk.

Here we propose a complementary platform, the Nano-membrane Experiment for Space-time Tremors (NEST), where the levitated disk is replaced by an optically trapped nanomechanical membrane resonator 
\cite{Fdisk,ni2012enhancement,clark2024optically}.
Millimeter-scale membranes can be reliably fabricated \cite{chakram2014dissipation,chowdhury2025optomechanical} and integrated into optical cavities via their supporting chip
\cite{thompson2008strong,rossi2018measurement}, thereby removing the primary technical obstacles of levitated geometries.
Owing to the resonant detection principle, which relies on the membrane's ultrahigh $Q$-factor, the instantaneous detection bandwidth is narrow, favoring continuous signals such as those from rapidly rotating neutron stars or axion superradiance.

A distinctive feature of NEST is its dual sensitivity. 
In addition to HFGWs, the detector can probe certain models of vector dark matter (VDM), 
such as the dark photon \cite{PhysRevD.84.103501,arias2012wispy,Vector-DM,VectorWaveDM,PRLLevitatedDM}. 
In our configuration, differential acceleration arises from the VDM's baryon–lepton ($B-L$) coupling with the silicon membrane and a germanium reference mass attached to the chip, whose distinct neutron-to-nucleon ratios generate a material-dependent force \cite{graham2016dark,carney2021ultralight}. 
Thus, both gravitational waves and vector dark matter drive the same mechanical mode, allowing a unified search strategy within a single device. 
The projected sensitivity is competitive with levitated-sensor proposals for HFGWs \cite{Nanodisk2022} and, in a best-case scenario, surpasses existing bounds from Eöt-Wash and LIGO/Virgo in the kHz band for $B-L$–coupled dark matter \cite{VDMLimitsEotWash,VDMLimitsLIGO}.

\begin{figure}
\centering
  {\includegraphics[width=0.47\textwidth]{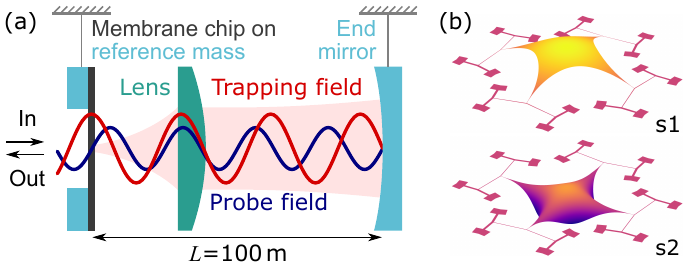}}   
   \caption{
   (a) Schematic of the NEST optomechanical detector platform. 
   A nanomechanical silicon (Si) membrane is used as the input mirror of an optical cavity of length $L$. 
   An optical trapping field with power $\Ptrap$ both tunes the membrane’s effective resonance frequency $\Omega_{\mathrm{eff}}$, and drives membrane motion when gravitational waves modulate the cavity length, which is amplified by the membrane at its resonance frequency.
   The amplified membrane motion is read out with a probe field. 
   The distinct baryon–lepton composition of the Si membrane and the Ge reference mass additionally enables sensitivity to $B-L$–coupled vector dark matter. 
   The reference mass, the lens, and the end mirror are suspended. The full detector consists of two such cavities forming the arms of a Michelson interferometer for common-mode noise suppression, as in Advanced LIGO \cite{Adv-LIGO-Specs} and the levitated-sensor detector proposal \cite{Nanodisk2022}. 
   (b) Mode shapes of the first (s1) and second (s2) symmetric trampoline modes used in the sensitivity analysis.
   \label{fig:general}}
\end{figure}

Figure~\ref{fig:general}(a) illustrates the NEST detector concept. 
A nanomechanical membrane, attached to a $B-L$ reference mass via its supporting chip, is used as the input mirror of an optical cavity, while also being trapped at the antinode of an optical trapping field (red).
When a passing GW changes the cavity length, the node of the trapping field remains anchored at the end mirror even as it moves due to its higher reflectivity and lack of interference from input light \cite{AggarwalEndMirror}, displacing the antinode relative to the membrane it is trapping and therefore actuating it.
If the GW frequency matches the membrane's effective resonance frequency, $\omega_{\mathrm{GW}} = \Omega_{\mathrm{eff}}$, this motion is resonantly amplified by the effective quality factor $Q_\mathrm{eff}=\Omega_\mathrm{eff}/\gammatot$, with $\gammatot$ the total damping rate [Eq.~(\ref{eq:gammas})] \cite{Nanodisk2013,Nanodisk2022}. 
For vector dark matter, an oscillating background field produces a spatially uniform differential acceleration between the membrane and the reference mass it is attached to if they are made of different materials and if their separation is small compared to the dark matter de Broglie wavelength. 
This acceleration is also resonantly enhanced when $\omega_{\mathrm{DM}} = \Omega_{\mathrm{eff}}$
\cite{carney2021ultralight,Vector-DM,chowdhury2025optomechanical}.
The membrane motion is read out with a much weaker probe field (blue) at a different longitudinal cavity mode, chosen such that its linear coupling to the membrane displacement is maximal, whereas that of the trapping field vanishes.
As the membrane forms the input mirror, its displacement changes the cavity length and thus phase-modulates the reflected probe field, generating signal sidebands as in the arm cavities of conventional interferometric detectors.
As the beam waist needs to be small at the membrane, a lens in the cavity is necessary to maintain cavity stability and to allow for feasible end mirror sizes.
The end mirror and the reference mass carrying the membrane will be suspended in a similar manner to the NEMO experiment \cite{NEMO,NEMOSuspension}; for the lens, a simpler suspension suffices, since its displacement couples to the cavity length only through residual reflections of its antireflection-coated surfaces.
More details on this configuration are available in Appendix~\ref{sec:overview}.
Common-mode technical noise such as vibrations, laser intensity fluctuations, or frequency noise, can couple to the gravitational-wave readout as sidebands on the laser carrier frequency.
By having two such cavities form the arms of a symmetric Michelson interferometer, common-mode rejection ensures these sidebands are reflected back to the input port and do not affect the detection port.
The gravitational wave modulates the two arms with opposite sign, such that the signal sidebands appear at the detection port, where they are read out by homodyne detection against a local oscillator, as in Advanced LIGO \cite{Adv-LIGO-Specs,capote2025advanced}. 

The effective resonance frequency can be tuned by changing the trapping field power via the optical spring effect \cite{ni2012enhancement}:
\begin{equation}
    \Omega_\mathrm{eff}=\sqrt{\Omega_\mathrm{m}^2+k_\mathrm{opt}/\meff}.
    \label{eq:omega_tune}
\end{equation}
Here, $\Omega_\mathrm{m}/2\pi$ and $\meff$ are the membrane's intrinsic resonance frequency and effective mass, and $k_\mathrm{opt}=\left(16\pi/\lambda\right)\left(|r|/|t|\right)\left(\Ptrap/c\right)$ is the optical spring constant, with $\lambda$ and $\Ptrap$ the trapping field wavelength and power, and $r$ and $t$ the membrane’s reflection and transmission coefficients \cite{jayich2008dispersive}.

For the membrane geometry we consider high-$Q$ trampoline resonators with branched tethers \cite{reinhardt2016ultralow,norte2016mechanical,Trampoline} as their designs allow for low intrinsic dissipation and are compatible with optical tuning.
Their first two symmetric modes (s1 and s2) are shown in Fig.~\ref{fig:general}(b). 

Similar membranes have already been fabricated from silicon nitride at the $\SI{}{mm}$-scale with $Q$-factors up to \SI{2e8}{} at room temperature \cite{Trampoline}.
Reaching the mechanical $Q$-factors on the order of $10^{10}$-$10^{11}$ required for sensitivity can be realized by scaling up the membranes to the \SI{}{cm}-scale and using crystalline silicon at cryogenic temperatures, which we simulate to form the basis of the sensitivity projections presented below.

\section{Model Setup}
\label{sec:model-setup}
The detector sensitivity to both high-frequency gravitational waves (Eq.~\ref{eq:HFGW}) and vector dark matter (Eq.~\ref{eq:VDM}) scales with the ratio
\begin{equation}
\sqrt{S_F}/m_{\mathrm{eff}},
\end{equation}
where $S_F$ is the noise spectral density.
For our sensitivity calculations, we take into account membrane thermal noise, shot noise, back-action noise, and thermal noise of the other cavity components, i.e., lens, end mirror, and reference mass (see Appendix~\ref{sec:noises}).
At resonance, the noise is completely dominated by the membrane thermal noise
\begin{equation}
S_F = 4 k_\mathrm{B} T m_{\mathrm{eff}} \gammatot
\label{eq:therm_noise}
\end{equation}
with temperature $T$ \cite{saulson1990thermal}. 
The total mechanical damping rate
\begin{equation}
\gammatot = \gammag + \gammasc/N_i + \gammam
\label{eq:gammas}
\end{equation}
includes contributions from residual gas damping $\gamma_g$, photon-recoil damping $\gamma_{\mathrm{sc}}$, and intrinsic mechanical loss $\gamma_m$. 
Here $N_i = k_B T/\hbar \Omega_{\mathrm{eff}}$ is the mean thermal occupation number.

Gas damping in the molecular flow regime is given by \cite{GasPressureSensor}
\begin{equation}
\gammag = \frac{P}{\rho d} \sqrt{\frac{32 M_{\mathrm{gas}}}{\pi R T}},
\end{equation}
with membrane density $\rho$ and thickness $d$, as well as gas molar mass $M_{\mathrm{gas}}$ and pressure $P$.

Photon scattering from the membrane leads to recoil damping \cite{Fdisk}
\begin{equation}
\gammasc = \frac{\lambda \Omega_{\mathrm{eff}}}{16 (n^2-1) d \mathcal{F}_{\mathrm{mem}}}
\left( \frac{w}{a} \right)^2,
\end{equation}
where $n$ is the refractive index, $\lambda$ the laser wavelength, $w$ the beam waist, and $a$ an effective pad radius taken as the radius of the largest circle inscribed in the membrane pad (see Fig.~\ref{fig:general}(b)).
The membrane-limited finesse $\mathcal{F}_{\mathrm{mem}}$ depends sensitively on $w/a$; we fix $w/a=0.6$, which balances scattering loss and mode distortion yielding $\mathcal{F}_{\mathrm{mem}}\approx 5.5\times 10^3$ \cite{Fdisk,ni2012enhancement,muller2015enhanced}.
We consider $w \gtrsim 375~\mu\mathrm{m}$; an intracavity lens keeps the beam diameter below \SI{21}{mm} along the entire cavity, so that the beam passes well within the $\approx\SI{50}{mm}$ free aperture of the ALPS~II beam pipe \cite{albrecht2020straightening}, while the cm-scale optics are placed outside the magnet string (Appendix~\ref{sec:overview}).

The intrinsic mechanical dissipation is given by 
\begin{equation}
\gammam = \Omega_{\mathrm{eff}}/Q,
\end{equation}
where the quality factor $Q$ is obtained from finite-element simulations via $Q = 2\pi W/\Delta W$ \cite{yu2012control,tsaturyan2017ultracoherent,muller2015enhanced,ni2012enhancement}. 
The elastic energy $W$ and bending-related loss $\Delta W$ follow standard thin-plate expressions; mode shapes $u(x,y)$ and effective masses $m_{\mathrm{eff}}$ are computed using COMSOL Multiphysics (v6.2). 




\subsection{Operating Environment and Optical Cavity}
We assume cryogenic operation at $T=4\,\mathrm{K}$ and pressure $P=10^{-11}\,\mathrm{torr}$, where hydrogen is the dominant residual gas species \cite{H2CryoVacuum1,H2CryoVacuum2}.
These conditions suppress thermal noise and minimize gas damping. They are within reach of the DESY ALPS~II infrastructure, which already operates an ultrahigh-vacuum beam pipe of more than \SI{100}{m} length inside superconducting magnet strings cooled with supercritical helium at \SI{4.5}{K}, and where the full experiment is intended to be located \cite{Brotherton2025ALPS,singh2026design,albrecht2020straightening}.
Only the immediate vicinity of the membrane needs to reach $10^{-11}$ \si{torr}, which can be achieved locally by cryopumping at \SI{4}{K} combined with differential pumping \cite{H2CryoVacuum1,H2CryoVacuum2}.

The operating wavelength is $\lambda=1550~\mathrm{nm}$, providing minuscule optical loss in the silicon membrane \cite{cole2013tenfold,cole2016high,PhysRevLett.117.213604}.
We use an optical cavity of length $L = 100~\mathrm{m}$ and finesse $\mathcal{F}_\mathrm{cav} \sim 10$, set by the membrane reflectivity, such that the cavity bandwidth exceeds the targeted 0.5–40\,kHz range, allowing the intracavity field to adiabatically follow length modulations.
We define $H\left(\Omega\right)=\sqrt{1+\left(2\Fcavity/\pi\right)^2\sin^2\left(\Omega L/c\right)}$ as the cavity response function \cite{Nanodisk2022,Nanodisk2013}. 
\subsection{Membrane Parameters}
The membrane material is crystalline silicon (Si), chosen for its ultra-low optical and mechanical loss~\cite{Qint-1p2e5,Qint-RT-cryo}, and high thermal conductivity at cryogenic temperatures~\cite{asheghi1997phonon,robinson2019crystalline}. 
Assuming surface loss to be the dominant dissipation channel~\cite{villanueva2014evidence,Qint-1p2e5,Si-rho-Y-Qint}, we scale the intrinsic quality factor linearly with membrane thickness $d$ as $Q_\mathrm{int} = 3.5\times10^5\, d / 120\,\mathrm{nm}$, consistent with measurements on oxide-free membranes~\cite{Qint-1p2e5}. 
Material parameters used in the simulations are listed in Table~\ref{tab:parameters}.  

Membrane geometries (overall extent, pad size, thickness, and internal stress) are optimized via parameter sweeps to minimize $\sqrt{S_F}/\meff$ at each target frequency. 
The pad size is constrained by optical scattering through the minimum allowable $\mathcal{F}_\mathrm{mem}$ and by the membrane geometry and supporting tethers. 
The overall lateral dimensions are limited to $8\times8~\mathrm{cm}^2$, compatible with fabrication on a six-inch wafer.


\subsection{Optically-Tuned Membrane Designs}
\label{sec:individual-designs}
For a first practical implementation we consider six membranes of thickness \SI{100}{nm}, that can be placed in the cavity one-at-a-time for dedicated runs with optical trapping powers up to \SI{20}{W}. 
\SI{20}{W} was chosen as a tentative upper range as existing devices in the literature have already demonstrated optical tuning ability up to at least \SI{17}{W} \cite{ni2012enhancement}, and we can confirm via COMSOL simulations that this power does not significantly increase membrane temperature.
Motivated by demonstrated devices \cite{Si-rho-Y-Qint,ChristophHighStress2}, we assume an intrinsic stress of \SI{1.5}{GPa}.
Resonance frequencies are increased primarily by reducing pad size at fixed overall extent until the limit $w/a=0.6$ is reached; further increases are obtained by raising the stress to \SI{3}{GPa} via strain engineering, as demonstrated in SiN \cite{ChristophHighStress1}.

Together these designs span 0.5–\SI{18}{kHz}.
Optical spring tuning extends the s1-mode frequency by a factor of \mbox{$\sim1.5$–$1.9$} at \SI{20}{W}, while the s2 mode shifts by \mbox{$\sim1.03$–$1.06$}.

\subsection{Considered Best-Case Membrane Designs}
\label{sec:best-case}
To determine the limiting best-case scenario of the platform, i.e., an upper bound on its performance, we assume a membrane thickness of \SI{500}{nm} and an internal stress up to the theoretical strength limit of silicon, \SI{20}{GPa} \cite{Si-StrengthLimit}. 

Starting from the lowest frequency (largest extent and pad size), higher resonances for the s1 and s2 modes [Fig.~\ref{fig:general}(b)] are obtained by increasing the stress while keeping the overall geometry fixed. 
Once the stress limit is reached, further frequency increases are achieved by reducing the pad size or membrane extent. The maximum frequency is obtained for the constraint $w/a=0.6$.

\section{Sensitivity to High Frequency Gravitational Waves}
The strain sensitivity $h_\mathrm{min}$ at the membrane's resonance frequency $\omega_\mathrm{GW}=\Omega_\mathrm{eff}$ is given by the following equation \cite{Nanodisk2013,Nanodisk2022}:
\begin{equation}
    h_\mathrm{min}=\frac{2}{\omega_\mathrm{GW}^2L}\frac{\sqrt{S_F}}{\meff}H\left(\omega_\mathrm{GW}\right),
    \label{eq:HFGW}
\end{equation}

The red traces in \refSensitivity show the sensitivity of the six optically tuned membrane designs, i.e., how resonance frequency and sensitivity change as the trapping power is varied from 1 to \SI{20}{W}.
The blue traces correspond to the best-case scenario, where dark and light traces correspond to the s1 and s2 modes respectively.

The minimum design strain to which aLIGO is sensitive is shown in grey \cite{LIGOGWLimits}, which increases with frequency due to shot noise until it no longer has sensitivity.
This contrasts with the membrane-based setup whose minimum sensitive strain decreases with frequency due to the underlying principle of resonant force detection. 

Over most of the band, gas damping is the dominant loss mechanism; intrinsic mechanical loss dominates in parts of the range, and photon-recoil damping becomes relevant only toward the upper end of the best-case band.
For the six optically tuned designs, the sensitivity improves as the trapping power raises the resonance frequency.
For the s1 modes, the mechanical dissipation exceeds the gas damping by up to a factor of a few at the highest powers, so that the improvement with frequency begins to saturate: since the beam waist is much smaller than the pad to suppress scattering loss, the optical spring deforms the mode shape and thereby increases bending losses \cite{ni2012enhancement,muller2015enhanced}.
The s2 modes, whose frequencies shift by only 3--6\%, are deformed much less and remain close to the gas-damping limit up to \SI{20}{W}.


The individual damping contributions are shown in Fig.~\ref{fig:sensitivityandgamma}.

\begin{figure}[h]
\centering
\includegraphics[width=0.47\textwidth]{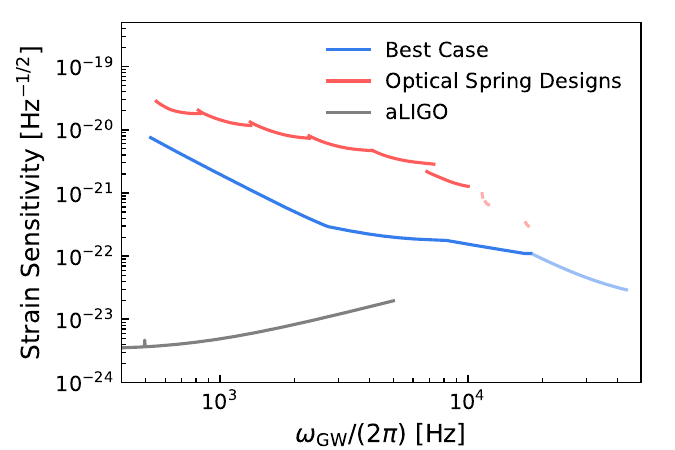}
   \caption{
   Strain sensitivity to high-frequency gravitational waves. 
   Best-case sensitivities (blue), together with six optically tuned membrane designs (red), where the resonance frequency is varied by increasing the trapping power from 1 to \SI{20}{W}. 
   Dark (light) shades represent the s1 (s2) mode.
   The aLIGO design sensitivity is shown for comparison \cite{LIGOGWLimits}. 
   \label{fig:sensitivity}
   }
\end{figure}

\section{Sensitivity to Vector Dark Matter}
Peak sensitivity to a dark matter mass $\mDM=\omegaDM\hbar/c^2$ occurs at resonance, $\omegaDM=\Omega_\mathrm{eff}$. 
The minimum detectable coupling at a signal-to-noise ratio of unity, averaged over vector polarizations, is given by \cite{Vector-DM,carney2021ultralight}:
\begin{eqnarray}
    g_\mathrm{min}=\frac{\sqrt{3/2}}{\beta \Delta_{12} a_0}
    \sqrt{\frac{\omega_\mathrm{DM}}{Q_\mathrm{DM}}}
    \frac{\sqrt{S_F}}{m_\mathrm{eff}}H\left(\omega_\mathrm{DM}\right) \nonumber \\
    \times
    \begin{cases}
    \left(\frac{\tau_\mathrm{DM}}{\tau}\right)^{1/2}, & \tau\lesssim\tau_\mathrm{DM}\\
    \left(\frac{\tau_\mathrm{DM}}{\tau}\right)^{1/4}, & \tau\gg\tau_\mathrm{DM}
    \end{cases}.
    \label{eq:VDM}
\end{eqnarray}

Here, $\Delta_{12}$ is a material-dependent factor encoding the differential acceleration between membrane and the reference mass it is attached to. For $B\!-\!L$–coupled dark matter,
$\Delta_{12}=\left|\frac{Z_1}{A_1}-\frac{Z_2}{A_2}\right|$, 
while for $B$ coupling,
$\Delta_{12}=\left|\frac{A_1}{\mu_1}-\frac{A_2}{\mu_2}\right|$, 
where $Z$ is the proton number, $A$ the nucleon number, $\mu$ the atomic mass, and indices 1 (2) denote membrane (reference mass).

The total measurement time is $\tau$, and $\tau_{\mathrm{DM}}=2Q_{\mathrm{DM}}/\omega_{\mathrm{DM}}$ is the dark-matter coherence time, with $Q_{\mathrm{DM}}\sim10^6$ reflecting the narrow linewidth of ultralight dark matter ($\mDM\lesssim\SI{1}{eV}/c^2$). As $\tau_{\mathrm{DM}}$ is only seconds to minutes in our band, interruptions between data stretches of several days do not degrade the sensitivity beyond the reduction of the accumulated $\tau$. The spatial overlap factor $
\beta=\frac{\iint u(x,y)\,\mathrm{d}A}{\iint u^2(x,y)\,\mathrm{d}A}
$
accounts for the coupling of a membrane mode to the assumed uniform VDM field, and $a_0=\SI{3.7e11}{m/s^2}$ sets the characteristic acceleration scale.

We use the same membrane geometries and material parameters as in the gravitational-wave analysis, considering a Si membrane attached to a Ge reference mass (atomic data in the Supplemental Material~\cite{SM}).

Figure~\ref{fig:VDM} shows the projected sensitivity to $B\!-\!L$ and $B$–coupled dark matter for the six optically tuned membranes (red) and best-case scenario (blue). Solid lines assume one year of integration, while dotted lines correspond to $\tau=\tau_{\mathrm{DM}}$. Dark (light) shades denote s1 (s2) modes.

\begin{figure}[h]
\centering
\includegraphics[width=0.47\textwidth]{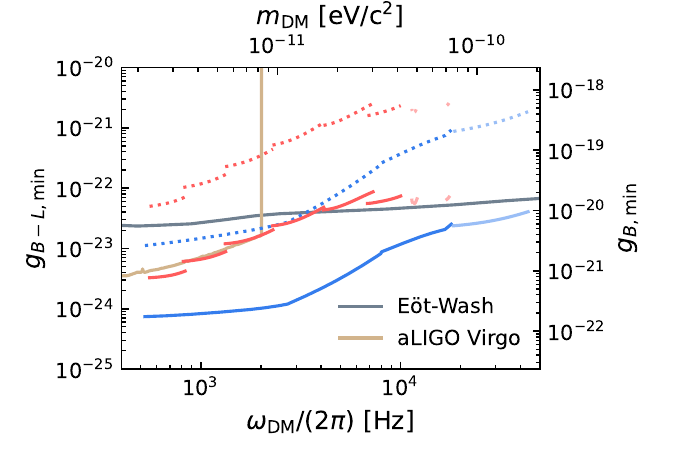}
   \caption{
   Sensitivity to $B-L$– and $B$–coupled vector dark matter at resonance. 
   Red curves correspond to the six optically tuned membrane designs; blue curves show the best-case scenario. 
   Dark (light) shades indicate s1 (s2) modes. 
   Solid lines assume an integration time of one year; dotted lines correspond to $\tau=\tau_{\mathrm{DM}}$. Existing limits from E\"ot-Wash and LIGO/Virgo are shown for comparison \cite{VDMLimitsEotWash,VDMLimitsLIGO}.}
    \label{fig:VDM}
\end{figure}

\section{Discussion}
We have presented NEST, an optomechanical platform for detecting high-frequency gravitational waves and ultralight vector dark matter.
In a limiting best-case scenario, it covers 0.5–40~kHz, corresponding to dark-matter masses of $2\times10^{-12}$–$2\times10^{-10}~\mathrm{eV}/c^{2}$, with a peak strain sensitivity of $3\times10^{-23}/\sqrt{\mathrm{Hz}}$ at \SI{40}{kHz}.
Toward this limit, six membranes with distinct intrinsic resonances and optical-spring tuning up to \SI{20}{W} are proposed as a first implementation within a Michelson configuration of two \SI{100}{m} cavities.

The projected gravitational-wave sensitivity is comparable to that of optically levitated nanodisk proposals in the overlapping 10–40~kHz band~\cite{Nanodisk2022} while avoiding the need for stable levitation.
An intracavity lens keeps the beam diameter below \SI{21}{mm} over the full \SI{100}{m}, so that the beam passes through the ALPS~II beam pipe, with cm-scale optics placed outside the magnet string.
The same lens-relayed cavity could also ease the implementation of the levitated-sensor detector~\cite{Nanodisk2022}, whose current design relies on a meter-scale end mirror, by reducing the required mirror size to the centimeter scale.

For vector dark matter, the best-case scenario surpasses existing bounds from Eöt-Wash and LIGO/Virgo across $2\times10^{-12}$–$2\times10^{-10}~\mathrm{eV}/c^{2}$ for one year of integration, opening a new window for physics beyond the Standard Model.

Existing infrastructure such as ALPS~II at DESY, which already operates a cryogenic ultrahigh-vacuum beam pipe of more than \SI{100}{m} length, provides a realistic path toward implementation~\cite{Brotherton2025ALPS,singh2026design}, with the stringent pressure of \SI{1e-11}{torr} required only in the immediate vicinity of the membrane.
The dimensions of the NEMO suspension setup also fit comfortably within the tunnel constraints \cite{NEMOSuspension}.

Future extensions may push the accessible frequency range toward the hundreds of kilohertz using alternative membrane geometries such as phononic-crystal designs~\cite{tsaturyan2017ultracoherent,rossi2018measurement}. 
Enhanced optical trapping via engineered reflectivity \cite{fan2002analysis,lousse2004angular,bernard2016precision,enzian2023phononically} or cavity-mode hybridization could further increase tunability \cite{AntiCrossingNature,muller2015enhanced}, in particular for the s2 modes, which are not yet limited by trapping-induced loss at \SI{20}{W}, while multi-membrane configurations \cite{Groeblacher} may strengthen optomechanical coupling and enable simultaneous multi-frequency searches.

\begin{acknowledgments}
We acknowledge funding awarded by a PIER Seed Project - project number PIF-2025-01.
We thank Aaron Spector, Albert Schliesser, Andreas Ringwald, Axel Lindner, Diego Blas, Krisztian Peters, and Nancy Aggarwal for helpful discussions. 
\end{acknowledgments}

\nocite{AtomicAbundances,AtomicWeights2,Cerdonio2001,Cole2013,Cole2023,ColeCryogenic2012,Craig2019,CryogenicMaterialModel,Didio2026,Franc2009,Glassbrenner1964,Haggerty2014,Hong2013,Komma2012,McGuigan1978,Mende2026,Nawrodt2011,Passler2013,Sparks1967,WortmanEvans1965,corruccini1960specific,shen2015optimized,shilov2025temperature}
\bibliography{apssamp}

\section*{End Matter}
\appendix

\section{Parameters and damping contributions}
\label{sec:params}
Table~\ref{tab:parameters} lists the baseline parameters used in the simulations.
The mechanical quality factor is obtained from finite-element simulations of the mode shapes as $Q=2\pi W/\Delta W$, with $W$ the stored elastic energy and $\Delta W$ the bending-related energy loss per cycle~\cite{yu2012control,tsaturyan2017ultracoherent}; details are given in the Supplemental Material~\cite{SM}.
Figure~\ref{fig:sensitivityandgamma} shows the resulting damping contributions entering Eq.~(\ref{eq:gammas}), together with the effective mass.

\begin{table}[!htb]
\centering
\caption{Baseline parameters used in simulations. Material constants are for crystalline silicon, where values given in parentheses represent the limiting best-case scenario.}
\label{tab:parameters}
\begin{tabular}{lcc}
\hline\hline
Parameter & Symbol & Value \\
\hline
Young’s modulus \cite{Si-rho-Y-Qint}& $Y$ & 169~GPa \\
Mass density \cite{Si-rho-Y-Qint}& $\rho$ & 2330~kg/m$^3$ \\
Poisson ratio \cite{SI-Poisson}& $\nu$ & 0.278 \\
Refractive index (1550 nm) \cite{Si-indexofrefraction}& $n$ & 3.45 \\
Thickness & $d$ & \SI{100}{nm} (\SI{500}{nm})\\
Internal stress & $\sigma$ & 1.5--\SI{3}{GPa} (\SI{20}{GPa}) \\
Temperature & $T$ & 4~K \\
Gas pressure (H$_2$) & $P$ & $10^{-11}$~torr \\
Molar mass (H$_2$) \cite{H2-molarmass} & $M_\mathrm{gas}$ & 2.016~g/mol \\
Cavity length & $L$ & 100~m \\
Cavity finesse & $\mathcal{F}_\mathrm{cav}$ & 10 \\
Laser wavelength & $\lambda$ & 1550~nm \\
Membrane-limited finesse & $\mathcal{F}_\mathrm{mem}$ & $5.5\times10^3$ \\
Beam waist to pad radius ratio & $w/a$ & 0.6 \\
\hline\hline
\end{tabular}
\end{table}

\begin{figure}[!htb]
\centering
\includegraphics[width=0.47\textwidth]{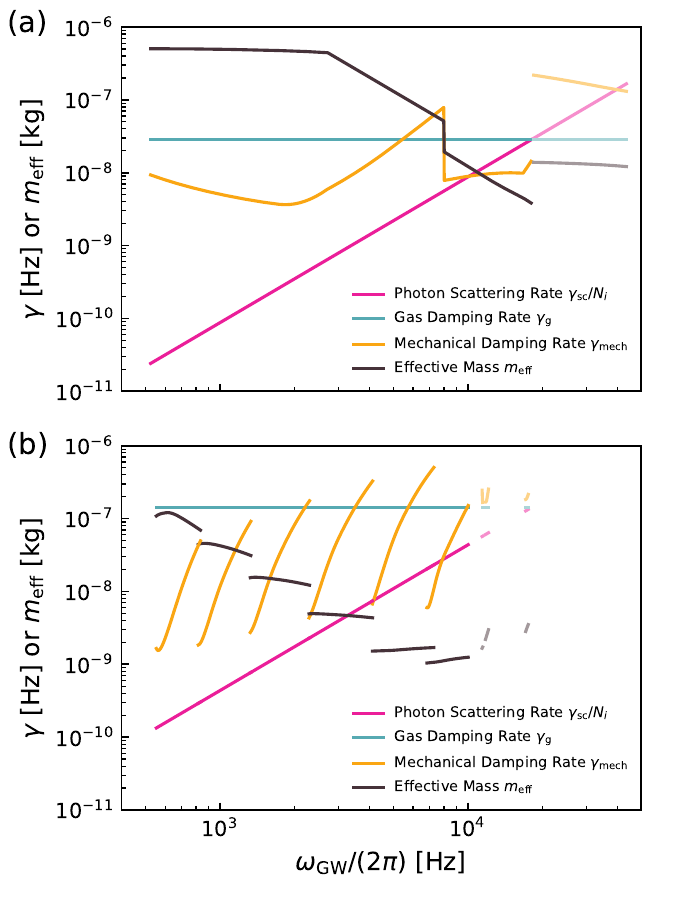}
   \caption{
   Individual damping contributions entering Eqs.~\ref{eq:gammas} together with the effective mass $\meff$. 
   (a) Best-case membrane designs. 
   (b) Six optically tuned membrane designs. 
   Shown are photon-scattering damping $\gamma_{\mathrm{sc}}/N_i$, gas damping $\gamma_g$, intrinsic mechanical damping $\gamma_m$, and the effective mass. Discontinuities arise from transitions between membrane geometries that optimize the sensitivity at different frequencies.
   Dark (light) shades indicate the s1 (s2) mode. 
}
    \label{fig:sensitivityandgamma}
    
\end{figure}

\section{Noise sources}
\label{sec:noises}
Besides membrane thermal noise, Eq.~(\ref{eq:therm_noise}), we consider probe back-action, shot noise, and thermal noise of the other cavity components.
Force noises on the membrane convert to strain via Eq.~(\ref{eq:HFGW}).
The probe back-action force noise is $S_{F,\mathrm{ba}}=64\hbar\Fcavity^2P_\mathrm{in}/(\pi c\lambda)$~\cite{Vector-DM}, kept small by a low probe input power $P_\mathrm{in}=\SI{10}{\mu W}$; the trapping field adds no linear back-action, as the membrane sits at its antinode, where the linear optomechanical coupling vanishes.
Shot noise is a displacement readout noise, $S_{d}=\pi\hbar c\lambda/(64\Fcavity^2P_\mathrm{in})$~\cite{Vector-DM}, which converts to strain through the mechanical susceptibility $\chi$ as $h_\mathrm{min}=2H\sqrt{S_d}/(L\,\meff\Omega_\mathrm{eff}^2|\chi|)$.
Membrane thermal noise dominates and exceeds shot noise by $R\gtrsim10^{6}$ in amplitude (Fig.~S1).
Owing to this margin, fluctuations of the trapping power, which shift $\Omega_\mathrm{eff}$ via the optical spring, leave the sensitivity unaffected as long as the shifts remain within $\sim\gammatot R/2$; this requires a relative intensity noise of the trapping field of $\lesssim10^{-5}$, within reach of established laser power stabilization~\cite{Kwee2012,Li2025intensity}.
The noise-limited strains of all sources and their derivation are given in the Supplemental Material~\cite{SM}.

\begin{figure}[!htb]
\centering
\includegraphics[width=0.36\textwidth]{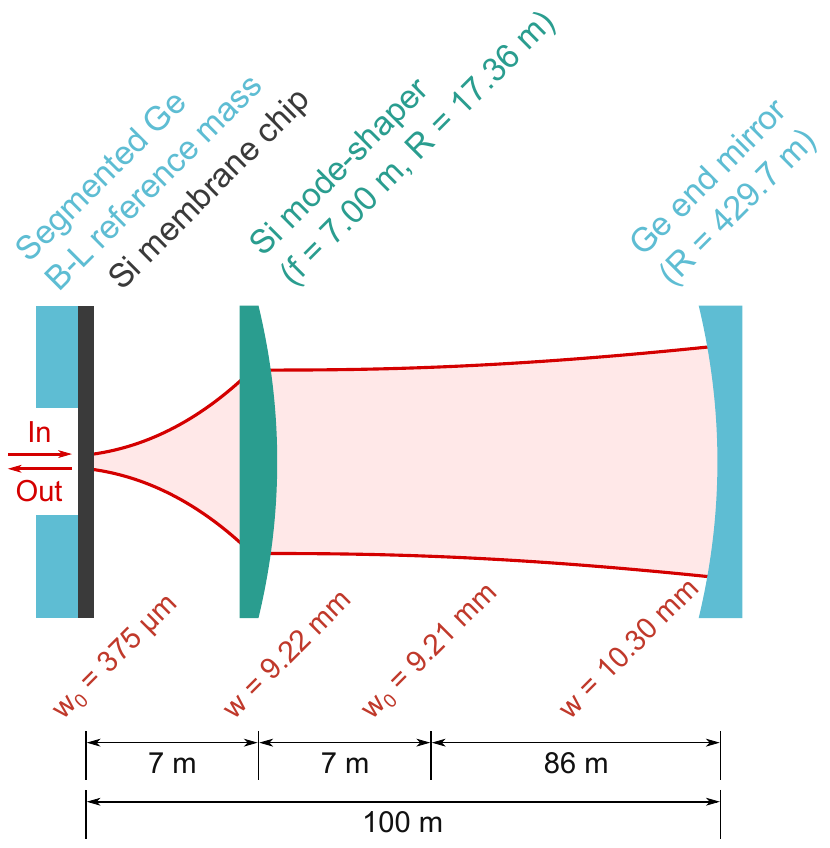}
\caption{Layout of a NEST cavity; two of them are arranged in a Michelson configuration. Material and coating parameters of the cavity components are given in the Supplemental Material~\cite{SM}.}
\label{fig:NEST_layout}
\end{figure}

\section{Optical layout and thermal noise of the cavity components}
\label{sec:overview}

Figure~\ref{fig:NEST_layout} shows the layout of a NEST cavity. 
The operating temperature is $4$~K, with a residual gas pressure of $10^{-11}$~torr at the
membrane. The silicon membrane chip, carrying the germanium
$B\!-\!L$ reference mass, forms the near end of a $100$~m cavity and is
illuminated with a waist of $w_0 = 375\,\mu$m. Free propagation over the first
$7$~m expands the beam to $9.22$~mm, where a silicon mode-shaping lens
($f = 7.00$~m, $R = 17.36$~m) collimates it to a $9.21$~mm waist. The beam then
propagates $86$~m to a germanium end mirror ($R_3 = 429.7$~m), where it has
grown to $10.30$~mm. Placing the only intracavity optic where the beam has
already expanded by free propagation is what keeps its thermal-noise
contribution small: substrate Brownian noise scales as $1/\sqrt{w}$, coating
Brownian as $1/w$, and bulk thermorefractive noise as $1/w^{2}$. The resulting
cavity is comfortably stable, with $g_1 g_2 = 0.800$, and the mode field
diameter stays below $21$~mm throughout. 
Optical trapping of a comparable nanomechanical element in a table-top optical cavity has been demonstrated in Ref.~\cite{ni2012enhancement}.

\begin{figure}[!htb]
\centering
\includegraphics[width=0.47\textwidth]{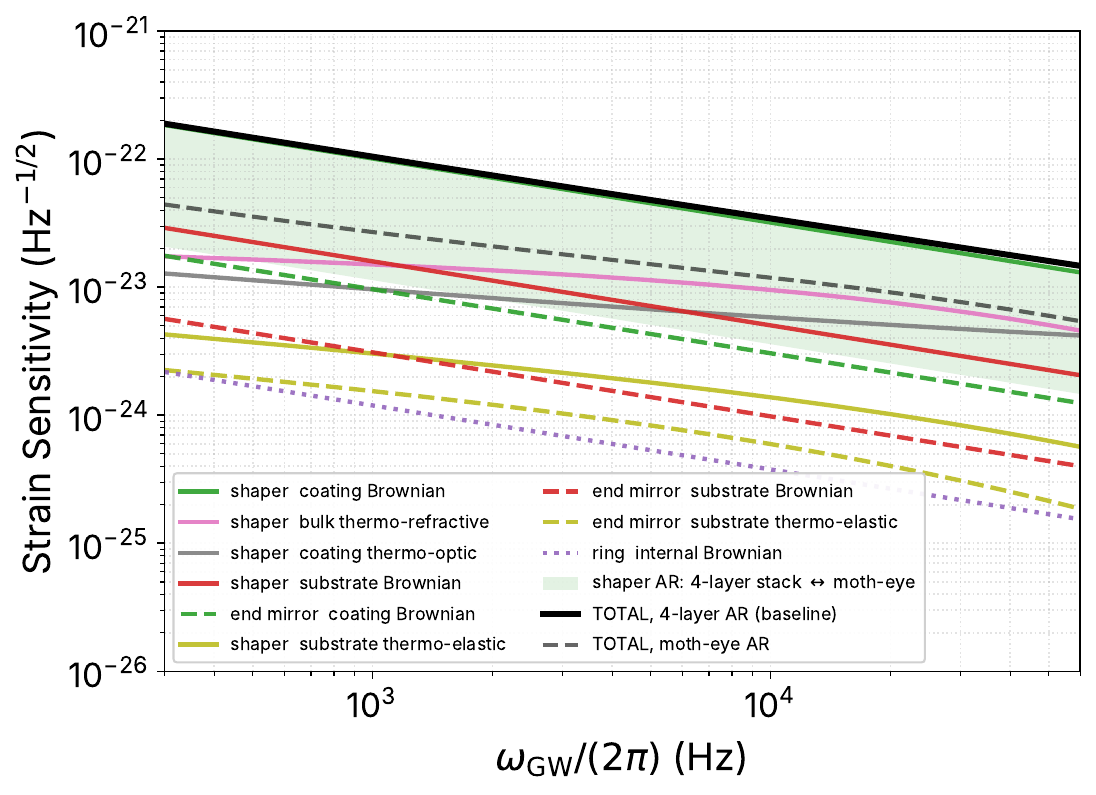}
\caption{Thermal noise budget of the cavity components other than the membrane (Fig.~\ref{fig:NEST_layout}). The membrane noise terms are discussed in Appendix~\ref{sec:noises}.}
\label{fig:noise_budget}
\end{figure}

Figure~\ref{fig:noise_budget} shows the thermal noise budget of the cavity components other than the membrane, computed with \textsc{pygwinc}~\cite{pygwinc}, the noise-budget tool of the LIGO Scientific Collaboration. 
Three of its default models are outside their range of validity for a cryogenic, small-beam, intracavity-transmissive configuration; the models used in their place, and one term for which \textsc{pygwinc} provides no model, are described in the Supplemental Material~\cite{SM}. 
Thermal fluctuations of these components change the optical path length between membrane and end mirror and thus displace the trapping-field antinode in the same way as a gravitational wave; an effective displacement noise $S_x$ therefore corresponds to a strain noise $h = 2\sqrt{S_x}/L$ per cavity, irrespective of the membrane response. For both the deposited four-layer antireflection stack of the lens (baseline) and an etched moth-eye structure, their thermal noise remains below the membrane-limited sensitivity $h_\mathrm{min}$ across the band (compare Figs.~\ref{fig:noise_budget} and~\ref{fig:sensitivity}), i.e., subdominant to the membrane thermal noise.

\end{document}


\title{Supplemental Material: An optomechanical platform for high-frequency gravitational wave and vector dark matter detection}
\author{David Rousso}
\affiliation{Deutsches Elektronen-Synchrotron DESY, 22603 Hamburg, Germany}
\author{Moritz Bjoern Kristiansson Kunze}
\affiliation{Deutsches Elektronen-Synchrotron DESY, 22603 Hamburg, Germany}
\author{Christoph Reinhardt}
\affiliation{Deutsches Elektronen-Synchrotron DESY, 22603 Hamburg, Germany}
\date{\today}
\maketitle

\section{Atomic data}
Table~\ref{tab:atomicweights} lists the atomic data entering the material-dependent factor $\Delta_{12}$ of the vector dark matter analysis.

\begin{table}[h]
\centering
\caption{Proton number, $Z$, nucleon number, $A$, and the atomic mass, $\mu$, of the membrane (Si) and reference-mass (Ge) materials used in the calculation of $\Delta_{12}$.}
\label{tab:atomicweights}
\begin{tabular}{lrllr}
\hline\hline
 & $Z$  & $A$  & $\mu$ \cite{AtomicWeights2} & Abundance \cite{AtomicAbundances} \\ \hline
Si   & 14 & 28 & 27.9769265         & 0.92255   \\
     &    & 29 & 28.9764947         & 0.04672   \\
     &    & 30 & 29.9737701         & 0.03073   \\
     \hline
Ge   & 32 & 70 & 69.9242485 & 0.205219 \\
     &    & 72 & 71.92207582 & 0.274515 \\
     &    & 73 & 72.92345895 & 0.07768 \\
     &    & 74 & 73.921177761 & 0.365212 \\
     &    & 76 & 75.921402725  & 0.075 \\
\hline\hline
\end{tabular}
\end{table}

\section{Mechanical quality factor}
The quality factor is calculated from COMSOL simulations as $Q=2\pi W/\Delta W $ \cite{yu2012control,tsaturyan2017ultracoherent,muller2015enhanced,ni2012enhancement}, where $W=\Omega_\mathrm{eff}^2  \rho d \iint  u^2 \mathrm{d} A /2$ is the energy stored in the mode of normalized shape $u=u(x,y)$. 
$\Delta W=\pi Y d^3 / \left[12(1-\nu^2)\Qint\right]\iint \xi\left(x,y\right) \mathrm{d} A$ is the mode's bending-related energy loss, with Young's modulus $Y$, Poisson ratio $\nu$, intrinsic material quality factor $\Qint$, where $\xi(x,y)$ is given by \cite{yu2012control} 
\begin{equation}
\resizebox{0.9\linewidth}{!}{$
\displaystyle
\xi\left(x,y\right)\!=\!\left(
\frac{\partial^2 u}{\partial x^2}
+
\frac{\partial^2 u}{\partial y^2}
\right)^2-2(1-\nu)\!\left[
\frac{\partial^2 u}{\partial x^2}
\frac{\partial^2 u}{\partial y^2}
-
\left(\frac{\partial^2 u}{\partial x\partial y}\right)^2
\right].
$}
\end{equation}

\section{Derivation of the noise-limited strain}
\label{sec:noisederiv}

For our sensitivity studies, we consider the noise sources of shot noise, probe back-action, membrane thermal noise, and coating/optics thermal noise. 
Membrane thermal noise dominates hence its focus in the main text.
Probe back-action noise, membrane thermal noise, and coating/optics thermal noise are all noises on the force on the membrane, whereas shot noise is a noise on the readout displacement.

The change in the cavity length $\Delta L$ is related to the gravitational wave strain $h$ via
\begin{equation}
    \Delta L=\frac{L}{2} h .
\end{equation}
The input displacement $x_{in}$ actuating the membrane is reduced by the cavity response $H(\omega_{GW})$:
\begin{equation}
    x_{in}=\frac{\Delta L}{H(\omega_{GW})} .
\end{equation}
The final membrane displacement is related to the input force $F_{in}$ via the susceptibility $\chi=\frac{1}{m_{\textrm{eff}}(\Omega_{\textrm{eff}}^2-\omega_{GW}^2+i\Omega_{\textrm{eff}}\omega_{GW}/Q_{\textrm{eff}})}$, and can be expressed in terms of a transfer function $T$ with respect to the input displacement. 
\begin{equation}
    x_{out}=\chi F_{in}=T\cdot x_{in}
\end{equation}
As the input force can be expressed in terms of the spring force, $F_{in}=k\cdot x_{in}=m_{\textrm{eff}}\Omega_{\textrm{eff}}^2x_{in}$, it follows that
\begin{equation}
    x_{out}=\chi m_{\textrm{eff}}\Omega_{\textrm{eff}}^2\cdot x_{in}
\end{equation}

If one has a noise on the input force to the membrane $F_{in}$, as from the above equations, the relation between the strain and the force is
\begin{equation}
    h=\frac{2}{L}\frac{1}{m_{\textrm{eff}}\Omega_{\textrm{eff}}^2}H(\omega_{GW})\cdot F_{in}
\end{equation}
and therefore
\begin{equation}
    h_{\textrm{min}}=\frac{2}{L}\frac{1}{m_{\textrm{eff}}\Omega_{\textrm{eff}}^2}H(\omega_{GW})\cdot \sqrt{S_F}
\end{equation}
where
\begin{equation}
    S_{F,\textrm{membrane-thermal}}=4k_B T m_{\textrm{eff}}\gamma_{tot}
\end{equation}
for the membrane thermal noise \cite{Nanodisk2022} [Eq.~(\ref{M-eq:therm_noise}) of the main text]
and 
\begin{equation}
    S_{F,\textrm{probe-backaction}}=\frac{64 \hbar \mathcal{F}_\mathrm{cav}^2 }{\pi  c \lambda}P_{\textrm{input,probe}}
\end{equation}
for the probe back-action noise \cite{Vector-DM}.
We set the probe input power to \SI{10}{\mu W}.
Note that we don't consider the back-action from the trapping beam as the linear dispersive coupling should vanish. 
The thermal noise of the other cavity components (lens, end mirror, and reference mass) is treated in Sec.~\ref{sec:departures}.

The shot noise is a readout noise on the final displacement $x_{out}$. 
This is related to the strain via
\begin{equation}
    h=\frac{2}{L}\frac{1}{m_{\textrm{eff}}\Omega_{\textrm{eff}}^2}\frac{1}{\chi}H(\omega_{GW})\cdot x_{out}
\end{equation}
and thus the noise-limited strain is
\begin{equation}
    h_{\textrm{min}}=\frac{2}{L}\frac{1}{m_{\textrm{eff}}\Omega_{\textrm{eff}}^2}\frac{1}{\chi}H(\omega_{GW})\cdot \sqrt{S_d}
\end{equation}
 where $S_d$ is the displacement noise, which for the shot noise takes the form
 \begin{equation}
     S_{d,\textrm{shot-noise}}
     =\frac{\pi\hbar c \lambda}{64 \mathcal{F}_{cav}^2 P_{\textrm{input, probe}}}
 \end{equation}
where the power here refers to the probe \cite{Vector-DM}.

The input and circulating powers are related by $P_{\textrm{input}}=\frac{\pi}{2\mathcal{F}_{cav}}P_{\textrm{circ}}$.

\begin{figure}[h]
\centering
\includegraphics[width=0.47\textwidth]{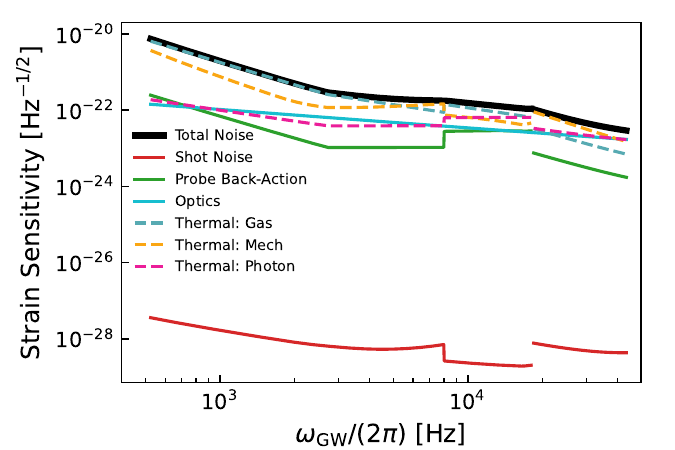}
\includegraphics[width=0.47\textwidth]{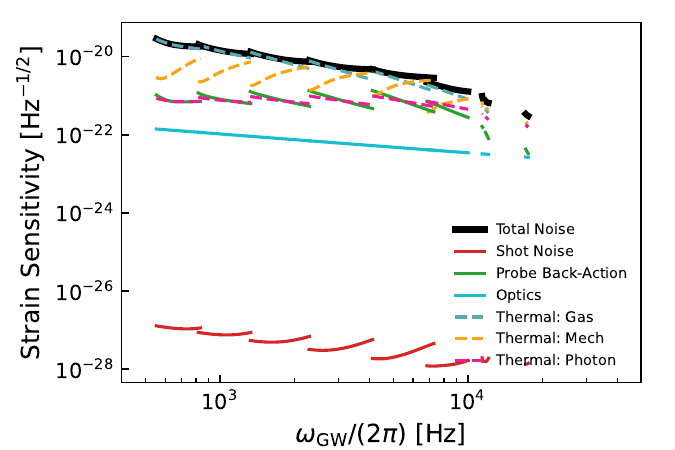}
\caption{Noise-limited strain for the considered noise sources, for (top) the best-case scenario and (bottom) the six optically tuned membrane designs.}
\label{fig:noisesSM}
\end{figure}

Figure~\ref{fig:noisesSM} compares the noise-limited strains of the considered noise sources.

\section{Departures from the default \textsc{pygwinc} treatment}
\label{sec:departures}

This section describes the deviations from the default \cite{pygwinc} configuration. 
For the baseline four-layer AR stack, the corrected budget is lower than the stock-default result by a roughly flat factor of $21$--$24$ across the band, because the stock-default path invokes the singular reflective-stack coating formula on a near-unity-transmission surface (Sec.~\ref{sec:artreatment}) rather than a frequency-dependent effect.
For the alternative moth-eye structure, the gap is instead a factor of $1.5$ at $60\,$kHz, rising to $\approx35$ at $300\,$Hz, driven by the non-adiabatic thermorefractive treatment of Sec.~\ref{sec:thermorefractive}. A reader reproducing this budget with \textsc{pygwinc} defaults will therefore obtain a substantially different curve regardless of which AR treatment is used; the flags controlling each departure are exposed at the top of the accompanying script.

\subsection{Non-adiabatic substrate thermoelastic noise}
\label{sec:cerdonio}

\textsc{pygwinc} implements the Braginsky--Gorodetsky--Vyatchanin 
result, Eq.~(5) of Ref.~\cite{Nawrodt2011},
\begin{equation}
  S^{\rm bulk}_{\rm TE}(f) = \frac{4k_{\rm B}T^{2}\alpha^{2}(1+\sigma)^{2}\kappa}
       {\sqrt{\pi^{5}}\,\rho^{2}C^{2}w^{3}f^{2}} ,
  \label{eq:bgv}
\end{equation}
valid only in the adiabatic regime, i.e.\ where the thermal diffusion length
\begin{equation}
  \ell_{\rm th} = \sqrt{\kappa/(\rho C\,\omega)}
  \label{eq:lth}
\end{equation}
is small compared with the beam radius $w$.  At $4\,$K the heat capacity has
collapsed while $\kappa$ remains large, so the thermal diffusivity
$D=\kappa/\rho C$ is of order $10^{5}$ times its room-temperature value and the
condition is violated across the entire detection band.  Reference
\cite{Nawrodt2011} states explicitly that this assumption breaks down for
crystalline materials at low temperature and that the true thermoelastic noise
is then \emph{smaller} than Eq.~(\ref{eq:bgv}) predicts.

We therefore use the general treatment of Cerdonio \emph{et al.}
\cite{Cerdonio2001}, Eqs.~(7)--(8) of Ref.~\cite{Nawrodt2011},
\begin{equation}
  S^{\rm bulk}_{\rm TE}(f) = \frac{4}{\sqrt{\pi}}\,
     \alpha^{2}(1+\sigma)^{2}\,\frac{k_{\rm B}T^{2}w}{\kappa}\;J[\Omega],
  \label{eq:cerdonio}
\end{equation}
\begin{equation}
  J[\Omega] = \sqrt{\frac{2}{\pi^{3}}}
     \int_{0}^{\infty}\!\!{\rm d}u \int_{-\infty}^{+\infty}\!\!{\rm d}v\;
     \frac{u^{3}e^{-u^{2}/2}}{(u^{2}+v^{2})\left[(u^{2}+v^{2})^{2}+\Omega^{2}\right]} .
  \label{eq:Jomega}
\end{equation}
The double integral is evaluated in polar coordinates $u=r\cos t$, $v=r\sin t$,
which removes the integrable singularity at the origin that makes the Cartesian
form converge poorly,
\begin{equation}
  J[\Omega] = \sqrt{\frac{2}{\pi^{3}}}
     \int_{0}^{\infty}\!\!{\rm d}r \int_{-\pi/2}^{\pi/2}\!\!{\rm d}t\;
     \frac{r^{2}\cos^{3}\!t\;e^{-r^{2}\cos^{2}\!t/2}}{r^{4}+\Omega^{2}} .
\end{equation}
Reference \cite{Nawrodt2011} does not define the normalisation of
$\Omega=\omega/\omega_{c}$.  We fix it by requiring that Eq.~(\ref{eq:cerdonio})
reduce to Eq.~(\ref{eq:bgv}) in the adiabatic limit, which gives
\begin{equation}
  \omega_{c} = \frac{2\kappa}{\rho C w^{2}},
  \qquad
  R(\Omega) \equiv \frac{S_{\rm exact}}{S_{\rm BGV}} = J[\Omega]\,\Omega^{2}
  \;\xrightarrow[\Omega\to\infty]{}\; 1 .
\end{equation}
$J[\Omega]$ is tabulated once on a logarithmic grid and interpolated, with an
interpolation error below $10^{-3}$.  The same non-adiabatic treatment was
applied to all low-temperature configurations in Ref.~\cite{Franc2009}.

A further consequence is worth noting.  Equation~(\ref{eq:bgv}) alone predicts
$S\propto\kappa$, so that reducing the thermal conductivity would reduce the
noise.  With the correction included $R\propto\kappa^{-1.15}$, and the product is
nearly independent of $\kappa$: the apparent benefit of a lower conductivity is
an artefact of extrapolating Eq.~(\ref{eq:bgv}) beyond its turnover near
$\Omega\sim1$.

\subsection{Non-adiabatic substrate thermorefractive noise}
\label{sec:thermorefractive}

By default \textsc{pygwinc} evaluates the bulk thermorefractive noise in the same
adiabatic approximation,
\begin{equation}
  S_{\rm TR}(f) \;=\; \frac{4H\beta^{2}\kappa\,k_{\rm B}T^{2}}
                {\pi\,r_{0}^{4}\,\omega^{2}(\rho C)^{2}},
  \qquad r_{0}=w/\sqrt{2},
  \label{eq:tradiab}
\end{equation}
with $\beta = {\rm d}n/{\rm d}T$ and $H$ the transmitted thickness.  We instead
use the exact expression of Braginsky \emph{et al.}, which \textsc{pygwinc}
provides but does not select by default and which is valid at all diffusion
regimes.

The difference is large because $\ell_{\rm th}\gg w$ at low frequency: heat
diffuses across the sampled volume many times per cycle and averages the
temperature over a region much larger than the beam.  Equation~(\ref{eq:tradiab})
contains no such saturation and gives an amplitude spectral density
$\propto f^{-1}$, whereas the exact result is nearly flat, $\propto f^{-0.13}$.
For the silicon mode-shaper the ratio between the two is $2.0$ at $40\,$kHz
and $\approx88$ at $300\,$Hz.

This term enters at all only because the mode-shaper lies \emph{inside} the
resonator.  In a conventional Fabry--Perot arm cavity the input-mirror substrate
is outside the resonant path and the circulating field never traverses it, which
is why bulk thermorefractive noise is neglected in Ref.~\cite{Nawrodt2011}
(their footnote~2).

\subsection{Thermal expansion of the SiO$_2$:HfO$_2$ layer}
\label{sec:sio2hfo2alpha}

Some materials shrink, rather than expand, as they warm up at low
temperature. This is a known feature of open, loosely connected crystal
structures -- vitreous SiO$_2$ is the textbook example, where flexible
corner-linked building blocks let the material's vibrations pull it inward as
heat is added. HfO$_2$ is built very differently: a dense, tightly packed
crystal structure with no such flexible linkages, much closer in character to
sapphire than to silica glass. Haggerty \emph{et al.}~\cite{Haggerty2014}
measured HfO$_2$'s thermal expansion directly along all crystal directions and
found ordinary behaviour -- it expands on warming, with no sign reversal, just
some directional variation. HfO$_2$ is therefore not expected to share SiO$_2$'s
anomalous shrinkage at low temperature.
 
Having already validated Debye/Gr\"uneisen scaling of $\alpha$ to a few percent
for sapphire (Sec.~\ref{sec:sio2hfo2kappa} discusses the analogous argument for
$\kappa$), we scale HfO$_2$'s own room-temperature value the same way:
\begin{align}
  \alpha_{\mathrm{HfO}_2}(4\,\mathrm{K}) &= \alpha_{\mathrm{HfO}_2}(300\,\mathrm{K})
     \times \frac{C(4\,\mathrm{K})}{C(300\,\mathrm{K})} \\
     &= 3.8\times10^{-6}\times1.8\times10^{-4} \\
     &= 6.8\times10^{-10}\,\mathrm{K}^{-1},
\end{align}
using $\alpha_{\mathrm{HfO}_2}(300\,\mathrm{K})=3.8\times10^{-6}\,\mathrm{K}^{-1}$
from Ref.~\cite{Franc2009} and the Debye heat-capacity ratio computed from
Ref.~\cite{Mende2026}'s Table II sound velocities and Debye temperatures. Mixed
with SiO$_2$'s own measured $\alpha(10\,\mathrm{K})=-2.5\times10^{-7}\,\mathrm{K}^{-1}$
at Craig \emph{et al.}'s $27\,\%$ SiO$_2$ doping level,
\begin{align}
  \alpha_\mathrm{mix} &= 0.27\times(-2.5\times10^{-7}) + 0.73\times(6.8\times10^{-10}) \\
     &= -6.7\times10^{-8}\,\mathrm{K}^{-1}.
\end{align}
The result is dominated by the $27\,\%$ SiO$_2$ minority phase by a factor
$135$, because SiO$_2$'s expansion barely suppresses at low
temperature while HfO$_2$'s is nearly Debye-suppressed to zero.

\subsection{Thermo-optic coefficient of the SiO$_2$:HfO$_2$ layer}
\label{sec:sio2hfo2dndt}

No comparable structural argument bounds $dn/dT$. Thermal expansion is a
purely anharmonic lattice property -- the Gr\"uneisen relation is defined by
it -- so it Gr\"uneisen-scales in any material that doesn't have SiO$_2$'s open, flexible structure.
$dn/dT$ additionally carries a direct electronic contribution from the
temperature dependence of the band gap, which need not track the lattice heat
capacity at all.

Shilov \emph{et al.}~\cite{shilov2025temperature} measured the optical band gap of
amorphous HfO$_2$ from $7$ to $296$~K and fit it to the standard Fan
electron-phonon model, extracting an effective phonon energy of $30\,\mathrm{meV}$.
Evaluating that model gives
\begin{equation}
  \frac{dE_g/dT\,(4\,\mathrm{K})}{dE_g/dT\,(300\,\mathrm{K})} \sim 1\times10^{-34},
\end{equation}
i.e.\ the band-gap-shift channel is suppressed exponentially (Bose--Einstein
phonon freeze-out) and is utterly negligible at $4$~K. This is a real,
quantified result, but it does not resolve the question: sapphire's
\emph{measured} cryogenic $dn/dT$ ($9\times10^{-8}$, Tomaru \emph{et al.}, via
Ref.~\cite{Franc2009}) is only $\sim150\times$ below its room-temperature
value, far short of an exponential suppression. Some other channel, not
captured by the single-phonon Fan picture, evidently dominates real cryogenic
$dn/dT$ in wide-gap oxides, and nothing here rules out an analogous channel in
HfO$_2$.

We therefore do not attempt a first-principles value. Instead we quote the
largest $dn/dT$ for which the reported budget changes by less than $3\,\%$
from the $dn/dT\to0$ case ($1.738\times10^{-23}\to1.790\times10^{-23}$,
$0.869\to0.895\,h_\mathrm{min}$):
\begin{equation}
  dn/dT = 3\times10^{-7}\,\mathrm{K}^{-1},
\end{equation}
a stated \emph{threshold}, not a measurement. For scale, this is $30\,\%$ of
SiO$_2$'s own measured cryogenic value ($1.01\times10^{-6}$ at $30$~K,
Ref.~\cite{Franc2009}). That comparator is itself likely an overestimate of a
true $4$~K value: SiO$_2$'s $dn/dT$ falls monotonically from
$\sim8\times10^{-6}$ at $300$~K to $1.01\times10^{-6}$ at $30$~K
(Ref.~\cite{Franc2009}, Tables 2 and 4); if the same decreasing trend
continues below $30$~K, the true $4$~K value is lower still, making our
threshold more conservative than the $30\,\%$ figure alone suggests.

\subsection{Coating Brownian noise on a transmitting face}
\label{sec:artreatment}

\textsc{pygwinc}'s \texttt{coating\_brownian} implements the formalism of Hong
\emph{et al.} \cite{Hong2013}.  That result contains two logically separate
ingredients: a \emph{mechanical} part, which propagates the bulk and shear noise
fields of each layer to a surface displacement using the full Poisson-ratio
transfer functions of their Table~I; and an \emph{optical weighting}, which
weights the layers by the standing-wave field distribution because the noise is
referred to the phase of the \emph{reflected} beam.  The optical weighting enters
through $\partial\log\rho/\partial\phi_{k}$, with $\rho$ the coating amplitude
reflectivity.

An antireflection coating is designed so that $\rho\rightarrow0$.  The optical
weighting is therefore \emph{singular}, not merely inaccurate, when applied to
one: our four-layer AR design has power transmission $T=0.99987$ and the routine
returns $2.4\times10^{-22}\,{\rm Hz}^{-1/2}$, whereas detuning the layer
thicknesses by $5\,\%$ ($T=0.971$) collapses the result to
$7.7\times10^{-24}\,{\rm Hz}^{-1/2}$ --- a factor $31$ for a $5\,\%$ change in
geometry.  We stress that detuning is not a remedy: for a transmitting face the
readout senses the transmitted optical path, so the reflected-phase noise is not
the sensed quantity at any value of $\rho$.

For transmitting faces we therefore retain Hong's mechanical treatment and omit
the optical weighting.  Setting the weighting factors to unity, the per-layer
coefficients of their Eq.~(94) collapse to a sum weighted by geometric thickness,
\begin{align}
  q_{Bk} &= \left(C_{B}+D_{B}\right)^{2}\,\frac{d_{k}}{\lambda_{k}}, \\
  q_{Sk} &= \left[\left(C_{SA}+D_{SA}\right)^{2}+D_{SB}^{2}\right]\frac{d_{k}}{\lambda_{k}},
\end{align}
with $d_{k}$ the geometric thickness of layer $k$, $\lambda_{k}$ the wavelength
inside it, and the Hong transfer functions
\begin{align}
  C_{B}  &= \sqrt{(1+\sigma_{k})/2}, \\
  C_{SA} &= \sqrt{1-2\sigma_{k}},\\
  D_{B}  &= \frac{(1-\sigma_{s}-2\sigma_{s}^{2})\,Y_{k}}{\sqrt{2(1+\sigma_{k})}\,Y_{s}}, \\
  D_{SA} &= -\frac{(1-\sigma_{s}-2\sigma_{s}^{2})\,Y_{k}}{2\sqrt{1-2\sigma_{k}}\,Y_{s}},\\
  D_{SB} &= \frac{\sqrt{3}\,(1-\sigma_{k})(1-\sigma_{s}-2\sigma_{s}^{2})\,Y_{k}}
                 {2\sqrt{1-2\sigma_{k}}\,(1+\sigma_{k})\,Y_{s}}, &&
\end{align}
where $Y_{k},\sigma_{k}$ are the layer and $Y_{s},\sigma_{s}$ the substrate
elastic constants.  This is finite for every stack, carries the full Poisson-ratio
dependence, and uses the same underlying formalism as the reflective case ---
the two differ only in whether the optical weighting is included.

\paragraph{Consistency check.}
Table~\ref{tab:coatcompare} compares the two \textsc{pygwinc} variants with the
small-Poisson-ratio expression of Harry \emph{et al.}, Eq.~(9) of
Ref.~\cite{Nawrodt2011}, at $40\,$kHz.

\begin{table}[h]
\caption{\label{tab:coatcompare}%
Coating Brownian noise at $40\,$kHz, expressed as equivalent strain, for the
three stacks in the budget.}
\begin{ruledtabular}
\begin{tabular}{lccc}
 & HR AlGaAs & AR 4-layer & Moth-eye \\
\hline
$T$                & $0.099$              & $0.99987$            & $0.696$              \\
Hong, full         & $3.69\times10^{-24}$ & $2.41\times10^{-22}$ & $5.45\times10^{-25}$ \\
Hong, mech.\ only  & $1.64\times10^{-24}$ & $8.23\times10^{-24}$ & $1.27\times10^{-24}$ \\
Harry Eq.~(9)      & $2.25\times10^{-24}$ & $9.10\times10^{-24}$ & $1.42\times10^{-24}$ \\
\end{tabular}
\end{ruledtabular}
\end{table}

For the high-reflectivity stack, where the optical weighting is physically
correct, the full Hong result exceeds Eq.~(9) by a factor $1.64$.  That factor
decomposes cleanly as
\begin{equation}
  \underbrace{0.73}_{\text{mechanical}}\times
  \underbrace{2.25}_{\text{optical weighting}} = 1.64 ,
\end{equation}
so the discrepancy is attributable to the optical weighting and not to the
Poisson-ratio terms that Eq.~(9) omits.  Compared like for like, mechanical
against mechanical, the two formalisms agree to $27\,\%$ for the HR stack and to
better than $11\,\%$ for both antireflection treatments, with Eq.~(9) the more
conservative in every case.  We use the full Hong result for the reflective end
mirror and the mechanical-only result for the transmitting faces.

The displacement of a transmitting face couples to the antinode position with
weight $(n-1)$ rather than unity, since the air path lost at the face is partly
compensated by the substrate path gained.

\subsection{Mechanical loss of an etched antireflection structure}
\label{sec:motheye}

\textsc{pygwinc} has no model for a subwavelength etched (``moth-eye'')
antireflection structure \cite{shen2015optimized}.  We estimate its mechanical loss with the surface-loss
formalism of Gretarsson \emph{et al.}, Eq.~(29) of Ref.~\cite{Nawrodt2011},
\begin{equation}
  \phi_{\rm surface} = \mu\,\alpha_{s}\,\frac{S}{V},
  \label{eq:surfaceloss}
\end{equation}
where $\alpha_{s}$ is a material-dependent surface-loss parameter with the
dimension of a length, of order picometres, and $\mu$ a geometric factor of order
unity to three.  For a structured layer of depth $d\simeq400\,$nm and period
$p\simeq300\,$nm the surface-to-volume ratio is $S/V\simeq4/p$, so that with
$\mu=3$ and $\alpha_{s}=0.5\,$pm
\begin{equation}
  \phi_{\rm me} \simeq 3\times(0.5\times10^{-12}\,{\rm m})
  \times\frac{4}{300\times10^{-9}\,{\rm m}} \simeq 2\times10^{-5}.
\end{equation}
The structured region is then treated as an effective-medium layer of that loss
angle, with a Young's modulus taken as half the substrate value (corresponding to
roughly $50\,\%$ fill) and a mean refractive index of $2$.  


\subsection{Thermal properties of the SiO$_2$:HfO$_2$ layer}
\label{sec:sio2hfo2kappa}


Mende \emph{et al.}~\cite{Mende2026} measure the through-plane thermal
conductivity of amorphous HfO$_2$/SiO$_2$ \emph{nanolaminates} (20 bilayers of
$10$~nm each) down to $30$~K, and separately tabulate the Debye parameters
(density, sound velocity, Debye temperature) of the amorphous constituents,
enabling a Cahill--Pohl minimum-thermal-conductivity ($\kappa_\mathrm{min}$)
calculation from first principles.

Reference~\cite{Mende2026} model the effective conductivity of a laminate as a 
series combination of interface and intrinsic resistance,
\begin{equation}
  \frac{1}{\kappa_\mathrm{eff}} = \frac{2N}{G\,t_\mathrm{nl}} + \frac{1}{\kappa_\mathrm{avg}},
\end{equation}
where $2N/t_\mathrm{nl}$ is the interface density and $\kappa_\mathrm{avg}$ the
intrinsic (bulk) harmonic-mean conductivity of the constituents. Their
measurement matches $\kappa_\mathrm{min}+\mathrm{DMM}$ almost exactly for their
$10$~nm-period stack, because closely spaced interfaces suppress the
long-wavelength phonon transport that otherwise lets bulk glass exceed
$\kappa_\mathrm{min}$ (their result for bulk a-SiO$_2$ exceeds $\kappa_\mathrm{min}$
by a factor $\sim3$ at $30$~K). Our AR layers are individually
$130$--$400$~nm thick -- about $28\times$ thicker than their $10$~nm layers --
giving an interface density $30\times$ lower. In their own model this makes
the interface term negligible for our geometry, so $\kappa_\mathrm{eff}\to
\kappa_\mathrm{avg}$, the intrinsic bulk value.

We therefore estimate $\kappa_\mathrm{avg}$ from the measured $\kappa$ of
a-SiO$_2$ as given in Table~\ref{tab:coatparams} and an a-HfO$_2$ estimate obtained by applying the same measured-to-$\kappa_\mathrm{min}$ excess ratio found for a-SiO$_2$ ($\approx20\times$ at $10$~K, from the Debye parameters of
Ref.~\cite{Mende2026}) to HfO$_2$'s $\kappa_\mathrm{min}(10\,\mathrm{K})$, giving
$\kappa_{\mathrm{HfO}_2}\approx0.19$~W\,m$^{-1}$K$^{-1}$ and a harmonic mean
$\kappa_\mathrm{avg}\approx0.17$~W\,m$^{-1}$K$^{-1}$. This value assumes 
a-HfO$_2$ shares a-SiO$_2$'s excess-over-$\kappa_\mathrm{min}$ factor, which is a proxy rather than a direct measurement, and is reported as such (status \emph{E}) in
Table~\ref{tab:coatparams}.

The volumetric heat capacity $C_V=322\,\mathrm{J\,m^{-3}K^{-1}}$ (status
\emph{E}, Table~\ref{tab:coatparams}) is obtained the same way, without the
excess-factor step: Ref.~\cite{Mende2026}'s Debye parameters give
$C_V(4\,\mathrm{K})=397$ and $119\,\mathrm{J\,m^{-3}K^{-1}}$ for a-HfO$_2$ and
a-SiO$_2$ respectively (Eq.~D4 of that reference), combined at the
$27\,\%$:$73\,\%$ SiO$_2$:HfO$_2$ doping level of Ref.~\cite{Craig2019}. Unlike
$\kappa$, $C_V$ is a purely harmonic-lattice quantity with no analogous
excess-over-Debye anomaly to correct for, so the Debye value is used directly.

\begin{table*}
\caption{\label{tab:params}%
Material and configuration parameters used in the optical-component thermal-noise
budget. All values are at the $T=4$~K operating point unless stated otherwise.
Status: \emph{M} = direct measurement at cryogenic temperature; \emph{D} = derived; \emph{E} = estimate, no cryogenic measurement available.}
\begin{ruledtabular}
\begin{tabular}{llllc}
Quantity & Symbol & Value & Source & Status \\
\hline
\multicolumn{5}{l}{\textbf{Operating point}}\\
Wavelength                & $\lambda$        & $1550$~nm                    & design & --- \\
Cavity length             & $L$              & $100$~m                      & design & --- \\
Operating temperature     & $T$              & $4$~K                        & design & --- \\
Circulating power         & $P_\mathrm{circ}$& $20$~W                       & design & --- \\
Residual gas pressure     & $p$              & $10^{-11}$~torr              & design & --- \\
Number of arms            & $N$              & 2 (Michelson)                & design & --- \\
\hline
\multicolumn{5}{l}{\textbf{Optical layout} (mode-shaper at $7$~m; see Fig.~\ref{M-fig:NEST_layout} of the main text)}\\
Membrane waist            & $w_0$            & $375~\mu$m                   & design & --- \\
Beam radius at shaper     & $w_\mathrm{sh}$  & $9.22$~mm                    & ABCD & D \\
Beam radius at end mirror & $w_\mathrm{em}$  & $10.30$~mm                    & ABCD & D \\
Mode shaper radius of curv.& $R_\mathrm{MS}$            & $17.4$~m                    & ABCD & D \\
End-mirror radius of curv.& $R_\mathrm{EM}$            & $429.7$~m                    & ABCD & D \\
Shaper thickness          & $t_\mathrm{sh}$  & $10$~mm                      & design & --- \\
Cavity stability          & $g_1 g_2$        & $0.800$                       & ABCD & D \\
\hline
\multicolumn{5}{l}{\textbf{Silicon} (mode-shaper substrate)}\\
Young's modulus           & $Y$              & $162.4$~GPa                  & \cite{Franc2009} Tab.~3 @ 10~K & M \\
Poisson ratio             & $\sigma$         & $0.2205$                     & \cite{Franc2009} Tab.~3        & M \\
Density                   & $\rho$           & $2331$~kg\,m$^{-3}$          & \cite{Franc2009} Tab.~3 @ 10~K & M \\
Refractive index          & $n$              & $3.48$ @ 1550~nm, 30~K       & \cite{Franc2009} Tab.~3        & M \\
Specific heat             & $C$              & $1.68\times10^{-2}$~J\,kg$^{-1}$K$^{-1}$ & \cite{corruccini1960specific}\footnote{Tabulated
   1--50~K; log-log interpolated. Validates the Debye $T^3$ form to 2\% at 4~K, but that form diverges above
   $\sim$10~K so interpolation is used throughout.} & M \\
Thermal conductivity      & $\kappa$         & $297$~W\,m$^{-1}$K$^{-1}$    & \cite{Franc2009} Tab.~3 @ 4~K & M \\
Thermal expansion         & $\alpha$         & $5.63\times10^{-11}$~K$^{-1}$& \cite{Sparks1967}\footnote{Measured
   from 1.89~K; the paper gives $\alpha=(0.880\pm0.016)\times10^{-12}\,T^{3}$ for $T<15.5$~K. Gr\"uneisen scaling
   from the 300~K value would give $6.09\times10^{-11}$, high by 8\%.} & M \\
Thermo-optic coefficient  & $dn/dT$          & $5.86\times10^{-9}$~K$^{-1}$ & \cite{Komma2012}\footnote{Measured
   at 1550~nm over 5--300~K: $1.0\times10^{-8}$ at 5~K, $1.8\times10^{-4}$ at 300~K. We interpolate as
   $dn/dT\propto T^{2.39}$ through these anchors. Unlike sapphire, silicon's $dn/dT$ Gr\"uneisen-scales
   and vanishes properly as $T\to0$.} & M \\
Bulk loss angle           & $\phi_\mathrm{bulk}$ & $1\times10^{-9}$         & \cite{Franc2009} Tab.~3 @ 10~K & M \\
\hline
\multicolumn{5}{l}{\textbf{Germanium} (end-mirror substrate and $B\!-\!L$ reference mass)}\\
Young's modulus           & $Y$              & $103$~GPa                    & \cite{WortmanEvans1965} & M \\
Poisson ratio             & $\sigma$         & $0.28$                       & \cite{WortmanEvans1965} & M \\
Density                   & $\rho$           & $5323$~kg\,m$^{-3}$          & standard & --- \\
Specific heat             & $C$              & $3.44\times10^{-2}$~J\,kg$^{-1}$K$^{-1}$ & \cite{corruccini1960specific}\footnote{Validates
   the Debye $T^3$ form to 5\% at 4~K.} & M \\
Thermal conductivity      & $\kappa$         & $550$~W\,m$^{-1}$K$^{-1}$    & \cite{Glassbrenner1964} & M \\
Thermal expansion         & $\alpha$         & $4.99\times10^{-10}$~K$^{-1}$& \cite{Sparks1967}\footnote{$\alpha=(7.80\pm0.15)\times10^{-12}\,T^{3}$
   for $T<12$~K. Gr\"uneisen scaling would give $6.07\times10^{-10}$, high by 22\%. Substituting the measured values for
   both Si and Ge changes the total budget by 0.1\%.} & M \\
Bulk loss angle           & $\phi_\mathrm{bulk}$ & $1\times10^{-9}$         & \cite{McGuigan1978} @ 4~K & M \\
$(B\!-\!L)/\mu$           &                  & $0.5594$                     & isotope-averaged & D \\
\hline
\multicolumn{5}{l}{\textbf{$B\!-\!L$ reference mass} (segmented Ge blocks bonded to the chip frame)}\\
Total mass                & $M$              & $0.4$~kg                     & design & --- \\
First internal resonance  & $f_1$            & $>50$~kHz                    & design constraint & --- \\
\hline
\multicolumn{5}{l}{\textbf{Optical absorption} (enters photothermal noise only)}\\
Shaper, bulk + AR         &                  & $20$~ppm                     & 10~ppm\,cm$^{-1}$ over 10~mm, double pass & --- \\
End mirror, HR coating    &                  & $1$~ppm                      & typical for crystalline coatings & --- \\
\end{tabular}
\end{ruledtabular}
\end{table*}

\begin{table*}
\caption{\label{tab:coatparams}%
Coating parameters, at $T=4$~K unless stated otherwise. Two antireflection
treatments are compared for the mode-shaper faces: an etched subwavelength
(``moth-eye'') structure, and a deposited four-layer SiO$_2$/SiO$_2$:HfO$_2$
stack. The end mirror carries a crystalline GaAs/Al$_{0.92}$Ga$_{0.08}$As
high-reflectivity stack (20 pairs) in both cases.}
\begin{ruledtabular}
\begin{tabular}{llllc}
Quantity & Symbol & Value & Source & Status \\
\hline
\multicolumn{5}{l}{\textbf{Antireflection coating -- moth-eye (alternative)}}\\
Etch depth                & $d$              & $400$~nm                      & design & --- \\
Effective index           & $n$              & $2.0$                         & effective medium, $\sim$50\% fill & E \\
Effective modulus         & $Y'$             & half of substrate $Y$         & effective medium & E \\
Surface-loss parameter    & $\alpha_s$       & $0.5$~pm                      & Gretarsson \emph{et al.} model, \cite{Nawrodt2011} Eq.~(29) & E \\
Loss angle                & $\phi$           & $2\times10^{-5}$              & derived above, see Sec.~\ref{sec:motheye} & E \\
\hline
\multicolumn{5}{l}{\textbf{Antireflection coating -- 4-layer deposited (baseline)}}\\
Total thickness           & $d$              & $1116$~nm, 4 layers           & design & --- \\
\hline
\multicolumn{2}{l}{SiO$_2$ (low index)} & & & \\
Young's modulus           & $Y'$             & $72$~GPa                      & \cite{Craig2019} Tab.~I & M \\
Poisson ratio             & $\sigma$         & $0.159$                       & \cite{Franc2009} Tab.~4 @ 10~K & M \\
Refractive index          & $n$              & $1.44$                        & \cite{Craig2019} Tab.~I & M \\
Loss angle (450~$^\circ$C)& $\phi$           & $8.5\times10^{-4}$            & \cite{Craig2019} Tab.~I @ 10~K & M \\
Loss angle (600~$^\circ$C)& $\phi$           & $5.0\times10^{-4}$            & \cite{Craig2019} Tab.~I @ 10~K & M \\
Thermal expansion         & $\alpha$         & $-2.5\times10^{-7}$~K$^{-1}$  & \cite{Franc2009} Tab.~4 @ 10~K (negative) & M \\
Thermo-optic coeff.       & $dn/dT$          & $1.01\times10^{-6}$~K$^{-1}$  & \cite{Franc2009} Tab.~4 @ 30~K & M \\
Specific heat             & $C$              & $4$~J\,kg$^{-1}$K$^{-1}$      & \cite{Franc2009} Tab.~4 @ 10~K (range 1--7) & M \\
Thermal conductivity      & $\kappa$         & $0.13$~W\,m$^{-1}$K$^{-1}$    & \cite{Franc2009} Tab.~4 @ 10~K & M \\
\hline
\multicolumn{2}{l}{SiO$_2$:HfO$_2$ (high index)} & & & \\
Young's modulus           & $Y'$             & $180$~GPa                     & \cite{Craig2019} Tab.~I & M \\
Poisson ratio             & $\sigma$         & $0.25$                        & estimate & E \\
Refractive index          & $n$              & $1.91$                        & \cite{Craig2019} Tab.~I & M \\
Loss angle                & $\phi$           & $3.8\pm0.3\times10^{-4}$      & \cite{Craig2019} Tab.~I @ 10~K & M \\
Thermal conductivity      & $\kappa$         & $0.17$~W\,m$^{-1}$K$^{-1}$    & derived, see Sec.~\ref{sec:sio2hfo2kappa} & E \\
Thermal expansion         & $\alpha$         & $-6.7\times10^{-8}$~K$^{-1}$  & derived, structural argument, see Sec.~\ref{sec:sio2hfo2alpha} & E \\
Thermo-optic coeff.       & $dn/dT$          & $3\times10^{-7}$~K$^{-1}$     & stated threshold ($<3\%$ budget change), see Sec.~\ref{sec:sio2hfo2dndt} & E \\
Volumetric heat capacity  & $C_V$            & $322$~J\,m$^{-3}$K$^{-1}$     & derived from Debye theory, see Sec.~\ref{sec:sio2hfo2kappa} & E \\
\hline
\multicolumn{5}{l}{\textbf{High-reflectivity coating -- GaAs/Al$_{0.92}$Ga$_{0.08}$As, 20 pairs (both AR treatments)}}\\
\multicolumn{2}{l}{GaAs} & & & \\
Young's modulus           & $Y'$             & $88$~GPa                      & \cite{CryogenicMaterialModel}, cross-checked with \cite{Cole2023} & E \\
Poisson ratio             & $\sigma$         & $0.312$                       & \cite{CryogenicMaterialModel} & E \\
Density                   & $\rho$           & $5317$~kg\,m$^{-3}$           & \cite{Cole2023} & M \\
Refractive index          & $n$              & $3.40$ @ 1550~nm              & \cite{CryogenicMaterialModel} & E \\
Specific heat             & $C$              & $0.043$~J\,kg$^{-1}$K$^{-1}$  & \cite{Passler2013} @ 4~K & M \\
Thermal conductivity      & $\kappa$         & $\sim500$~W\,m$^{-1}$K$^{-1}$ & \cite{CryogenicMaterialModel}; range $10^2$--$10^3$, sample-dependent & E \\
Thermal expansion         & $\alpha$         & $\sim5\times10^{-8}$~K$^{-1}$ & \cite{CryogenicMaterialModel}; range $10^{-8}$--$10^{-7}$ & E \\
Thermo-optic coeff.       & $dn/dT$          & $\sim7\times10^{-9}$~K$^{-1}$ & \cite{CryogenicMaterialModel}, $T^3$ extrapolation & E \\
\hline
\multicolumn{2}{l}{Al$_{0.92}$Ga$_{0.08}$As} & & & \\
Young's modulus           & $Y'$             & $85$~GPa                      & \cite{CryogenicMaterialModel}, cross-checked with \cite{Cole2023} & E \\
Poisson ratio             & $\sigma$         & $0.323$                       & \cite{CryogenicMaterialModel} & E \\
Density                   & $\rho$           & $3900$~kg\,m$^{-3}$           & \cite{Cole2023} & M \\
Refractive index          & $n$              & $2.92$ @ 1550~nm              & \cite{CryogenicMaterialModel} & E \\
Specific heat             & $C$              & $\sim0.075$~J\,kg$^{-1}$K$^{-1}$ & \cite{CryogenicMaterialModel}; range $0.05$--$0.1$ @ 4~K & E \\
Thermal conductivity      & $\kappa$         & $\sim50$~W\,m$^{-1}$K$^{-1}$  & \cite{CryogenicMaterialModel}; range $10$--$10^2$, sample-dependent & E \\
Thermal expansion         & $\alpha$         & $\sim5\times10^{-8}$~K$^{-1}$ & \cite{CryogenicMaterialModel}; range $10^{-8}$--$10^{-7}$ & E \\
Thermo-optic coeff.       & $dn/dT$          & $\sim4\times10^{-9}$~K$^{-1}$ & \cite{CryogenicMaterialModel}, $T^3$ extrapolation & E \\
\hline
\multicolumn{2}{l}{Loss angle (both layers, same value assumed)} & & & \\
$4$--$10$~K               & $\phi$           & $4.5\times10^{-6}$            & \cite{ColeCryogenic2012, Didio2026} & M \\
$20$~K (interpolated)     & $\phi$           & $5\times10^{-6}$              & \cite{ColeCryogenic2012, Didio2026} & M \\
$300$~K                   & $\phi$           & $2.5\times10^{-5}$            & \cite{Cole2013, Didio2026} & M \\
\end{tabular}
\end{ruledtabular}
\end{table*}

\bibliography{apssamp}